# The Murphy-Good plot: a better method of analysing field emission data

Richard G. Forbes

*University of Surrey, Advanced Technology Institute & Dept. of Electrical and Electronic Engineering, Guildford, Surrey, GU2 7XH, UK.*

*e-mail: r.forbes@trinity.cantab.net*

Measured field electron emission (FE) current-voltage $I_m(V_m)$ data are traditionally analysed via Fowler-Nordheim (FN) plots, as $\ln\{I_m/V_m^2\}$ vs $1/V_m$. These have been used since 1929, because in 1928 FN predicted they would be linear. In the 1950s, a mistake in FN's thinking was found. Corrected theory by Murphy and Good (MG) made theoretical FN plots slightly curved. This causes difficulties when attempting to extract precise values of emission characterization parameters from straight lines fitted to experimental FN plots. Improved mathematical understanding, from 2006 onwards, has now enabled a new FE data-plot form, the "Murphy-Good plot". This plot has the form $\ln\{I_m/V_m^{(2-\eta/6)}\}$ vs $1/V_m$, where "eta" ($\eta$) depends only on local work function. Modern ("21st century") MG theory predicts that a theoretical MG plot should be "almost exactly" straight. This makes precise extraction of well-defined characterization parameters from ideal $I_m(V_m)$ data much easier. This article gives the theory needed to extract characterization parameters from MG plots, setting it within the framework of wider difficulties in interpreting FE $I_m(V_m)$ data (among them, use of the "planar emission approximation"). Careful use of MG plots could also help remedy other problems in FE technological literature. It is argued MG plots should now supersede FN plots.

# 1. Background

Field electron emission (FE) occurs in many technological contexts, especially electron sources and electrical breakdown. A need exists for effective analysis of measured FE current-voltage [$I_m(V_m)$] data, to extract emission characterization parameters. These include: parameters that connect field to voltage; the field enhancement factors (FEFs) often used to characterize large-area field-electron emitters (LAFEs); and parameters relating to emission area and area efficiency (the latter being a measure of what fraction of emitter area is emitting significantly). This article proposes a simple, new precise method for FE $I_m(V_m)$ data analysis, and urges its widespread adoption. This method, the Murphy-Good (MG) plot, would replace the traditional Fowler-Nordheim (FN) plot. To develop MG-plot theory efficiently, some discussion and refinement of traditional FN theory is needed first.

FN plots were introduced by Stern et al. [1] in 1929. They have the form $\ln\{I_m/V_m^2\}$ vs $1/V_m$ (or equivalent using other physical variables), and are used because the original 1928 FN equation [2] implied that FN plots of experimental data should be straight lines, with characterization data derivable from the slope and intercept.

However, in 1953, Burgess, Kroemer & Houston (BKH) [3] found a mathematical mistake in 1928 theoretical work by Nordheim [4], and a related physical mistake in FN's thinking. FN had assumed [2] that image-force rounding could be disregarded, and treated the electron tunnelling barrier as exactly triangular. BKH showed that rounding was much more important than FN had thought and Nordheim had calculated, and that (for emitters modelled as planar) it is *necessary* to base analyses on planar image-rounded barriers [often now called "Schottky-Nordheim" (SN) barriers]. Corrected analysis inserted a "barrier form correction factor" into the exponent of the original 1928 FN equation, and led to much higher tunnelling probabilities (by a factor of more than 100). This correction factor is given by an appropriate value of a special mathematical function (SMF) v($x$) now known [5] to be a special solution of the Gauss Hypergeometric Differential Equation. The "Gauss variable" $x$ is the independent variable in this equation.

[The Nordheim parameter $y$ used in older FE discussions is given by $y = +x^{1/2}$, but its *mathematical* usage can now be recognized as *perverse*—when a function "$\mathfrak{F}$" is the solution of a differential equation, mathematics does not normally represent $\mathfrak{F}$ as a function of the square root of the independent variable in the equation. The use of $y$ (rather than $x$ [$=y^2$]) in FE literature is due to an unfortunate arbitrary choice (separate from the above mistake) made by Nordheim in his 1928 paper. Although $y$ is useful as a modelling parameter in some theoretical discussions, hindsight indicates that choosing to use $x$ [$=y^2$] in 1928 would have proved much better mathematics (and better for discussing FE $I_m(V_m)$ data).]

In 1956, Murphy and Good (MG) [6] used the BKH results to develop a revised FE equation. [See Ref. [7] for a treatment that uses the modern "International System of Quantities" (ISQ) [8].] The zero-temperature version of their equation is called here the *Murphy-Good (MG) equation*, and in practice applies well at room temperatures,

The MG equation gives the local emission current density (LECD) $J_L^{MG}$ in terms of the local work function $\phi$ and local barrier field $F_L$. Initially, it is clearest to use the linked form

$$J_L^{MG} = t_F^{-2} J_{kL}^{SN}, \qquad (1.1a)$$

$$J_{kL}^{SN} = a\phi^{-1}F^2 \exp[-v_F b\phi^{3/2}/F_L], \qquad (1.1b)$$

where $a$ [$\cong 1.541434$ μA eV V$^{-2}$] and $b$ [$\cong 6.830890$ eV$^{-3/2}$ V nm$^{-1}$] are universal constants [9], often called the *first* and *second Fowler-Nordheim constants*, $v_F$ is the value of v($x$) that applies to the "reference SN barrier" defined by $\phi$ and $F_L$, and $t_F$ is the reference-barrier value of a related SMF t($x$) [7]. $J_{kL}^{SN}$ is called the *kernel current density for the SN barrier*, and can be evaluated *precisely* when $\phi$ and $F_L$ are known.

The correction factor $v_F$ is field-dependent (see below). This causes theoretical FN plots predicted by the MG equation to be slightly curved, rather than straight. This in turn causes very significant problems of detail [10], and the need for related procedures [11] when attempts are made to give well-defined *precise* meanings to the slope and intercept of the straight line fitted to a FN plot of experimental data. This article shows how to eliminate these problems, by finding a plot form that the MG equation predicts to be "almost exactly" linear.

## 2. Some general issues affecting field emission current-voltage data analysis

In fact, three other major problems affect both FN-plot and MG-plot interpretation, and need discussion. The FN and MG derivations disregard the existence of atoms and model the emitter surface as smooth, planar and structureless. This *planar emission approximation* is unrealistic, but creating reliably better theory is very difficult, although there are some atomic-level treatments, e.g. Refs [12] and [13]. At present, when applying a current-density equation to real emitters, this weakness must be explicitly formalized. To recognize both this difficulty, and all other factors omitted in deriving eq. (1.1b), the present author has replaced $t_F^{-2}$ in eq. (1.1a) by an "uncertainty factor" $\lambda$ of unknown exact value (see ref. [14] for recent discussion). I now write $J_L^{EMG} = \lambda J_{kL}^{SN}$, and call the revised equation (and variants using other physical variables) the *extended Murphy-Good (EMG) equation*. Current thinking [15] is that (for a SN barrier) $\lambda$ probably lies somewhere in the range $0.005<\lambda<14$.

The total emission current ($I_e$) is found by integrating $J_L^{EMG}$ over the emitter surface and writing the result as first shown below, where $A_n^{EMG}$ is the *notional emission area* (as derived using the EMG equation):

$$I_\mathrm{e}^\mathrm{EMG}(F_\mathrm{C}) \;=\; \int J_\mathrm{L}^\mathrm{EMG}\, \mathrm{d}A \;=\; A_\mathrm{n}^\mathrm{EMG} J_\mathrm{C}^\mathrm{EMG} \;=\; A_\mathrm{n}^\mathrm{EMG} \lambda\, J_\mathrm{kC}^\mathrm{SN} \;\equiv\; A_\mathrm{f}^\mathrm{SN} J_\mathrm{kC}^\mathrm{SN}\,. \tag{2.1}$$

The subscript "C" denotes characteristic values taken at some characteristic location on the emitter surface (in modelling, nearly always the emitter apex).

The second form follows from $J_\mathrm{C}^\mathrm{EMG} = \lambda\, J_\mathrm{kC}^\mathrm{SN}$. Often, $\lambda$ and $A_\mathrm{n}^\mathrm{EMG}$ are both unknown. Equations with two unknown parameters are inconvenient, so these are combined into a single parameter $A_\mathrm{f}^\mathrm{SN}$ [$\equiv \lambda A_\mathrm{n}^\mathrm{EMG}$] called the *formal emission area* (for the SN barrier).

Combining these various relations, and assuming that measured current $I_\mathrm{m}$ equals emission current $I_\mathrm{e}^\mathrm{EMG}$, yields the following EMG-theory equation:

$$I_\mathrm{m}(F_\mathrm{C}) \;=\; A_\mathrm{f}^\mathrm{SN} J_\mathrm{kC}^\mathrm{SN} \;=\; A_\mathrm{f}^\mathrm{SN} a\phi^{-1} F_\mathrm{C}^2 \exp[-\mathrm{v}_\mathrm{F} b \phi^{3/2}/F_\mathrm{C}]\,. \tag{2.2}$$

It needs to be understood that, although this is not explicitly shown, the values of $\phi$, $\mathrm{v}_\mathrm{F}$, $\lambda$, $A_\mathrm{n}$, and $A_\mathrm{f}$ depend on the choice of location "C".

When applying this equation to experiments, and "thinking backwards", $I_\mathrm{m}(F_\mathrm{C})$ is a measured quantity, and $J_\mathrm{kC}^\mathrm{SN}$ can be calculated precisely (when $\phi$ and $F_\mathrm{C}$ are known). Thus, the *extracted* parameter $\{A_\mathrm{f}^\mathrm{SN}\}^\mathrm{extr}$ [$= I_\mathrm{m}(F_\mathrm{C})/J_\mathrm{kC}^\mathrm{SN}$] is, in principle, a well-defined parameter that depends on the barrier form, but not on $\lambda$: thus, the symbol $\{A_\mathrm{f}^\mathrm{SN}\}^\mathrm{extr}$ carries the barrier label, rather than an equation label.

In practice, it is nearly always the *formal* area that is initially extracted from a FN or MG plot. Issues of how formal area relates to the notional area in some specific emission equation, or to geometrical quantities relating closely to real emitters, are matters for separate discussion later, maybe in many years' time when good values for $\lambda$ are known.

A second major problem lies in determining the relationship between the characteristic barrier field $F_\mathrm{C}$ and the measured voltage $V_\mathrm{m}$. I now prefer to write

$$F_\mathrm{C} \;=\; V_\mathrm{m}/\zeta_\mathrm{C}\,, \tag{2.3}$$

where $\zeta_\mathrm{C}$ is the *characteristic voltage conversion length (VCL)*, for location "C". Except in special geometries, $\zeta_\mathrm{C}$ is not a physical length. Rather, $\zeta_\mathrm{C}$ is a system characterization parameter: low VCL means the emitter "turns on" at a relatively low voltage $V_\mathrm{m}$.

So-called *ideal* FE devices/systems have $I_\mathrm{m}(V_\mathrm{m})$ characteristics determined *only* by the emission process and the system geometry, with no "complications" (see below); in this case, the VCL $\zeta_\mathrm{C}$ is *constant*, and related parameters (such as characteristic FEFs) can be derived from extracted $\zeta_\mathrm{C}$-values (see below).

However, real devices/systems may have "complications", such as (amongst others) leakage current, series resistance in the measuring circuit, current dependence in FEFs, and space-charge effects. These may cause "non-ideality" whereby $\zeta_C$ ceases to be constant but becomes dependent on voltage and/or current. This in turn may modify the FN or MG plot slope or cause plot non-linearity. In such cases, conventional FN-plot analysis may be likely to generate spurious results for characterization parameters [16]. This will also be true for MG-plot analysis.

Additional research is urgently needed on how to analyse and model the FE $I_m(V_m)$ characteristics of non-ideal devices/systems, but it will likely be many years before comprehensive theory exists. Hence, at present, FN and MG plots provide reliable emission characterization *only* for ideal devices/systems. For FN plots there is a spreadsheet-based [16] "orthodoxy test" that can filter out non-ideal data sets; a version for MG plots will be described elsewhere.

A third major problem is the following. For ideal real emitters, even if one assumes the emitter radius is large enough for the SN barrier to be an adequate approximation for evaluating tunnelling probabilities, one expects that $A_f^{SN}$ would depend on emitter shape and applied voltage. However, the FN and MG plot theories are built using the planar emission approximation. In this approach $A_f^{SN}$ is treated as a constant, with the extracted value $\{A_f^{SN}\}^{extr}$ derived—with varying degrees of precision— from the slope and intercept of a straight line fitted to an experimental FN or MG plot.

Current understanding is that $\{A_f^{SN}\}^{extr}$ is actually some kind of effective average value of $[I_m/J_{kC}^{SN}]$, taken over the range of $F_C$-values used in the experiments. But detailed physical interpretation of $\{A_f^{SN}\}^{extr}$ is an issue separate from whether the extracted value is a useful characterization parameter (which it is considered to be). Values of $\{A_f^{SN}\}^{extr}$ are presumed particularly useful for LAFEs, when comparing the properties of different emitting materials or processing regimes. Thus, having a simple method of extracting a numerically well-defined value (from a particular set of ideal experimental data) is expected to be helpful.

For LAFEs, a more useful property is perhaps the *extracted formal area efficiency* $\{\alpha_f^{SN}\}^{extr}$ (for the SN barrier), defined by

$$\{\alpha_f^{SN}\}^{extr} \equiv \{A_f^{SN}\}^{extr} / A_M , \qquad (2.4)$$

where $A_M$ is the *LAFE macroscopic area or ("footprint")*. Few experimental values have been reported for $\{\alpha_f^{SN}\}^{extr}$. It is thought [17] to be very variable as between LAFEs, but perhaps to often lie in the vicinity of $10^{-7}$ to $10^{-4}$. Clearly, if—for some particular LAFE material—data analysis showed (for example) that only $10^{-5}$ % of the footprint area was actually emitting electrons, then this might indicate scope for practical improvements. This parameter looks potentially useful for technology development.

## 3. Theory of Murphy-Good plots

Given the above context, MG-plot theory can now be developed. This is most easily done using *scaled* parameters and equations, as follows. The *scaled (barrier) field f* (for a barrier of zero-field height $\phi$) is a dimensionless physical variable formally defined, using the *Schottky constant* $c_S$ [$\equiv (e^3/4\pi\varepsilon_0)^{1/2}$] [9], by

$$f \equiv c_S^2 \phi^{-2} F_L \cong [1.439\,965 \text{ eV}^2 \text{ (V/nm)}^{-1}] \, \phi^{-2} F_L \, . \tag{3.1}$$

For a SN barrier of zero-field height $\phi$ (the "reference SN barrier"), the criterion $f=1$ defines a *reference field* $F_R$ [$=c_S^{-2}\phi^2$] at which the barrier top is pulled down to the Fermi level. For this barrier, $f=F_L/F_R$, and hence $F_L = f F_R$. It can be shown from Ref. [6] (but, better, see arXiv:1801.08251v2) that $v_F = v(x=f)$.

Scaling parameters $\eta(\phi)$ and $\theta(\phi)$ are defined by

$$\eta(\phi) = bc_S^2 \phi^{-1/2}, \quad \theta(\phi) = ac_S^{-4}\phi^3 \, . \tag{3.2}$$

Substituting $F_C = f_C F_R = f_C \, c_S^{-2} \phi^2$ into eq. (2.2), and writing $v_F$ explicitly as $v(f_C)$, yields the *scaled equation*

$$I_m(f_C) = A_f^{SN} \theta(\phi) f_C^2 \exp[-\eta(\phi) \cdot v(f_C)/f_C] \, . \tag{3.3}$$

For simplicity, we now normally cease to show the dependence of $\eta$ and $\theta$ on $\phi$.

A key development [18], in 2006, was the discovery of a simple good approximation for $v(f)$:

$$v_F = v(f) \approx 1 - f + (1/6) f \ln f \, . \tag{3.4}$$

In $0 \leq f \leq 1$, $v_F$ takes values between $v(f=0) = 1$ and $v(f=1) = 0$. For eq. (9), in $0 \leq f \leq 1$, Ref. [7] found the maximum error in $v(f)$ as 0.0024 and the maximum percentage error as 0.33%. High-precision numerical formulae for $v(f)$, with maximum error $8 \times 10^{-10}$ in $0 \leq f \leq 1$, are also known [7].

Setting $f=f_C$ and substituting eq. (3.4) into eq. (3.3) leads, after some re-arrangement, to

$$I_m(f_C) \approx A_f^{SN} \theta \cdot \exp\eta \cdot f_C^\kappa \cdot \exp[-\eta/f_C], \tag{3.5a}$$

$$\kappa \equiv 2 - \eta/6 \, . \tag{3.5b}$$

For an ideal device/system, eq. (3.1) can be used to define, by $V_{mR} = F_R \zeta_C \; [= c_S^{-2} \phi^2 \zeta_C]$, a *reference measured-voltage* $V_{mR}$ at which, at location "C", the SN barrier-top is pulled down to the Fermi level. It follows that

$$f_C = F_C/F_R = (V_m/\zeta_C)/(V_{mR}/\zeta_C) = V_m/V_{mR}, \tag{3.6}$$

and that eq. (3.5) can be rewritten as

$$I_m(V_m) \approx \{A_f^{SN}(\theta \exp\eta) \cdot V_{mR}^{-\kappa}\} \cdot V_m^{\kappa} \cdot \exp[-\eta V_{mR}/V_m], \tag{3.7}$$

and then

$$\ln\{I_m/V_m^{\kappa}\} \approx \ln\{A_f^{SN}(\theta \exp\eta) \cdot V_{mR}^{-\kappa}\} - \eta V_{mR}/V_m. \tag{3.8}$$

This is an equation for a *theoretical Murphy-Good plot*.

Since $A_f^{SN}$ is being treated as constant, and all parameters on the right-hand side (except $V_m$) are constants, eq. (3.8) must be a *straight line* with slope $S_{MG}$ and intercept $\ln\{R_{MG}\}$ given by:

$$R_{MG} = A_f^{SN}(\theta \exp\eta) \cdot V_{mR}^{-\kappa}, \tag{3.9}$$

$$S_{MG} = -\eta V_{mR} = -b\phi^{3/2}\zeta_C. \tag{3.10}$$

The subscript "MG" indicates that these parameters "belong to" a theoretical MG plot. It further follows that

$$R_{MG} \cdot (|S_{MG}|)^{\kappa} = A_f^{SN} \cdot \theta \cdot \exp\eta \cdot \eta^{\kappa} = A_f^{SN} \cdot \theta\eta^2 \cdot \exp\eta \cdot \eta^{-\eta/6}. \tag{3.11}$$

From equations above, $\theta\eta^2 = ab^2\phi^2 \; [\cong (7.192492 \times 10^{-5} \text{ A nm}^{-2} \text{ eV}^{-2})\phi^2]$. Thus, if $S_{MG}$ and $\ln\{R_{MG}\}$ are identified with the slope $S_{MG}^{fit}$ and intercept $\ln\{R_{MG}^{fit}\}$ of a straight line fitted to an experimental MG plot, the extracted values of the VCL $\zeta_C$ and formal emission area $A_f^{SN}$ are:

$$\zeta_C^{extr} = -S_{MG}^{fit}/b\phi^{3/2} \tag{3.12}$$

$$\{A_f^{SN}\}^{extr} = \Lambda_{MG} \cdot R_{MG}^{fit} \cdot (|S_{MG}^{fit}|)^{\kappa}, \tag{3.13}$$

where the *area extraction parameter $\Lambda_{MG}$ (when using an MG plot)* is given by

$$\Lambda_{MG}(\phi) \equiv 1/[(ab^2\phi^2)\cdot \exp\eta \cdot \eta^{-\eta/6}]. \tag{3.14}$$

An extracted area-efficiency value can be obtained from eqns (2.4) and (3.13), and an extracted value of macroscopic FEF $\gamma_M$ from eq. (3.12) and the relation

$$\gamma_M^{extr} = d_M / \zeta_C^{extr}, \tag{3.15}$$

where $d_M$ is the system distance used to define the FEF and related macroscopic field $F_M$.

Since expression (3.14) depends only on $\phi$, a table of $\Lambda_{MG}(\phi)$-values is easily prepared with a spreadsheet. Some illustrative values are shown in Table 1. $\Lambda_{MG}(\phi)$ is only weakly dependent on $\phi$, so uncertainty in the true $\phi$-value should cause little error in the extracted value of formal emission area

Table 1. Typical values of $\Lambda_{MG}(\phi)$ and related parameters.

| $\phi$ (eV) | $h$ | $\exp\eta \cdot \eta^{-\eta/6}$ | $\Lambda_{MG}(\phi)$ (nm$^2$/A) |
|---|---|---|---|
| 3.50 | 5.2577 | 44.85 | 25.31 |
| 4.00 | 4.9181 | 37.06 | 23.45 |
| 4.50 | 4.6368 | 31.54 | 21.77 |
| 5.00 | 4.3989 | 27.46 | 20.25 |
| 5.50 | 4.1942 | 24.34 | 18.89 |

The consistency of the above approach has been tested by simulating a MG plot over the range $0.15 \leq f_C \leq 0.35$, using input values $\phi$=4.50 eV and $A_f^{SN}$ = 100 nm$^2$. Depending on which small part ($\Delta f$=0.01) of the plot was used to derive values of $R_{MG}^{fit}$ and $S_{MG}^{fit}$, the extracted value $\{A_f^{SN}\}^{extr}$ lay between 97.8 nm$^2$ and 100.3 nm$^2$, i.e. a consistency discrepancy of around 2.5%. (For a FN plot, the equivalent consistency discrepancy, using a standard constant value for $\Lambda_{FN}^{SN}$, is around a factor of 2.)

## 4. Discussion

The essential merit of the Murphy-Good plot is that the whole tiresome apparatus [10,11] of slope and intercept correction factors, fitting points and chord corrections (needed for high precision when a FN plot is used with the EMG or MG equations) has been swept away. At this stage of development (but see below), FE data analysis has been restored to something like the 1929 simplicity found in Stern et al.

Use of a FN plot and the so-called *elementary FN-type equation* [14] (which is a simplified version of the original 1928 FN equation) is, of course, slightly more straightforward, algebraically.

However, it is known (e.g., Ref. [11]) that error by a large factor—typically around 100—is involved when this approach is used to carry out emission area extraction. So the MG plot needs to be used.

The author's view is that using MG plots should benefit three groups of experimentalists who currently use FN plots (and will also benefit the subject as a whole). Those who already use FN-plot interpretation theory based on eq. (3.4) will no longer need to use slope and intercept correction factors, or equivalent. Those who already use the MG equation, but use formulae based ultimately on 1970s approximations for $v_F$, such as those of Spindt et al. [19] or Shrednik [20], will get slightly more precise results than before, and will not have to use approximation formulae whose true origin may not always be obvious. But the largest group of beneficiaries may be those who analyse FN plots by using the elementary FN-type equation. (This practice may be partly due to a mis-statement, in an influential 2004 paper [21], about the suitability of the elementary FN-type equation as an approximation.) For this group, for ideal devices/systems, the simple formulae provided here allow them to precisely extract (from an MG plot) information about three characterization parameters, rather than one: the VCL, the FEF and the formal area efficiency.

The formulae here envisage that researchers will use their raw $I_m(V_m)$ data to make $I_m(V_m)$ MG plots, and will then apply an orthodoxy test [16]—which must be passed if values for extracted (and related) characterization parameters are to be regarded as trustworthy. As indicated earlier, an orthodoxy test already exists for FN plots, and a modified version will be made available shortly for MG plots. Hopefully, this should help to reduce the incidence of spuriously high FEF values reported in the literature.

Using $I_m(V_m)$-type MG plots could also help eliminate the widespread but unfortunate literature practice of pre-converting $I_m(V_m)$ data to become $J_M(F_M^{app})$ data before making a FN plot, where $F_M^{app}$ is the apparent macroscopic field obtained from the pre-conversion equation, and $J_M$ is the *macroscopic (or LAFE-average) current density* defined by $J_M=I_m/A_M$. This pre-conversion is almost always carried out by using a plausible but often *defective* conversion equation (defective because it can be invalid for non-ideal devices/systems). This in turn has often led to defective FN plots and spurious results for characterization parameters.

Another feature of experimental FE literature is that papers sometimes use *macroscopic* current densities to show data or make FN plots, but state a formula for *local* current density in the text, without drawing attention to the difference. This practice creates un-discussed apparent discrepancies between theory and experiment, sometimes by a factor of $10^6$ or more. Such confusions would be reduced if, instead, FE papers gave an equation for measured current, either an $I_m(F_C)$ equation of form (3) above, or a related $I_m(V_m)$ equation.

Finally, it is needful to remember that all FN and MG plots implicitly involve the (unrealistic) planar emission approximation. The issues of how best to include emitter shape, when predicting FE $I_m(V_m)$ characteristics or analysing experimental FE $I_m(V_m)$ data, are topics of active research, impracticable to summarize here. At present, no consensus exists on how best to perform data analysis for non-planar emitters, and significant amounts of detailed further research seem needed. It may take several years or more to reach consensus, and many further years to develop fully correct theory. Expectation is that, in due course (some, or likely many, years away), we shall need to move on from MG plots, and may possibly need to consider other analysis techniques, such as multi-parameter numerical fitting, rather than new plot forms. However, until this happens, Murphy-Good plots should be used because they are straightforward, and appear to be a much better approach to data analysis than Fowler-Nordheim plots.

**Acknowledgments**

Research by Dr Eugeni O. Popov and colleagues at the Ioffe Institute in Saint-Petersburg have been a major stimulus for this work. Their numerical simulations (e.g., Ref. [22]) relating to carbon nanotubes have looked for the value of *k* in the empirical FE equation $I_m = CV_m^k \exp[-B/V_m]$ proposed [23] some years ago. My thinking about how to find *C*-values has led to this article. I thank Dr Popov for numerous e-mail discussions.

**References**

1. Stern TE, Gossling BS, Fowler RH. 1929 Further studies in the emission of electrons from cold metals. *Proc. R. Soc. Lond.* A **124**, 699-723.
2. Fowler RH, Nordheim L, 1928 Electron emission in intense electric fields. *Proc. R. Soc. Lond.* A **119**, 173-181.
3. Burgess RE, Kroemer H, Houston JM. 1953 Corrected values of Fowler-Nordheim field emission functions $v(y)$ and $s(y)$. *Phys. Rev.* **90**, 515.
4. Nordheim LW. 1928 The effect of the image force on the emission and reflection of electrons by metals. *Proc. R. Soc. Lond.* A **121**, 626-639.
5. Deane JHB, Forbes RG. 2008 The formal derivation of an exact series expansion for the Principal Schottky-Nordheim Barrier Function $v$, using the Gauss Hypergeometric Differential Equation . *J. Phys. A: Math. Theor.* **41**, 395301.
6. Murphy EL, Good RH. 1956 Thermionic emission, field emission and the transition region. *Phys. Rev.* **102**, 1464-1473.
7. Forbes RG, Deane JHB. 2007 Reformulation of the standard theory of Fowler-Nordheim tunnelling and cold field electron emission. *Proc. R. Soc. Lond.* A **463**, 2907-2927.

8. *International Standard ISO 80000-1:2009. Quantities and units—Part 1 General,* (ISO, Geneva).
9. Forbes RG, Deane JHB. 2011 Transmission coefficients for the exact triangular barrier: an exact general analytical theory that can replace Fowler & Nordheim's 1928 theory. *Proc. R. Soc. Lond.* A **467**, 2927-2947. See Electronic Supplementary Material for information about special universal constants used in field emission.
10. Forbes RG, Fischer A, Mousa MS. 2013 Improved approach to Fowler-Nordheim plot analysis. *J. Vac. Sci. Technol.* B **31**, 02B103.
11. Forbes RG, Deane JHB. 2017 Refinement of the extraction-parameter approach for deriving formal emission area from a Fowler-Nordheim plot. 30th International Vacuum Nanoelectronics Conf., Regensburg, July 2017. (doi:10.13140/RG.2.2.33297.74083)
12. Modinos A. 2001 Theoretical analysis of field emission data. *Solid-State Electronics* **45**, 809-816.
13. Lepetit B. 2017 Electronic field emission models beyond the Fowler-Nordheim one. *J. Appl. Phys.* **122**, 215105. (doi:10.1063/1.5009064)
14. Forbes RG, Deane JHB, Fischer A, Mousa MS. 2015 Fowler-Nordheim plot analysis: a progress report. *Jordan J. Phys.* **8**, 125-147. (Also see: arXiv:1504.01634v7.)
15. Forbes RG. 2018 Comparison of the Lepetit field emission current-density calculations with the Modinos-Forbes uncertainty limits. 31st International Vacuum Nanoelectronics Conf., Kyoto, July 2018. (doi:10.13140/RG.2.2.35893.73440/1)
16. Forbes RG. 2013 Development of a simple quantitative test for lack of field emission orthodoxy. *Proc. R. Soc. Lond.* A **469**, 20130271. (doi:10.1098/rspa.2013.027)
17. Forbes RG. 2009 Use of the concept "area efficiency of emission" in equations describing field emission from large area electron sources *J. Vac. Sci. Technol.* B **27**, 1200-1203. (doi: 10.1116/1.3137964)
18. Forbes RG. 2006 Simple good approximations for the special elliptic functions in standard Fowler-Nordheim tunnelling theory for a Schottky-Nordheim Barrier. *Appl. Phys. Lett.* **89**, 113122. (doi:10.1063/1.2354582)
19. Spindt CA, Brodie I, Humphrey L, Westerberg ER. 1976 Physical properties of thin-film field emission cathodes. *J. Appl. Phys.* **47**, 5248-5263.
20. Shrednik VN. 1974 "Theory of field emission", Chap. 6 in: Elinson MI (ed.) *"Unheated Cathodes"* ("Soviet Radio", Moscow, 1974) (In Russian). See eq. 6.10.
21. de Jonge N, Bonard J-M. 2004 Carbon nanotube electron sources and applications, *Phil. Trans. R. Soc. Lond.* A **362**, 2239-2266. (doi:10.1098/rsta.2004.1438)
22. Popov EO, Kolosko AG, Filippov SV. 2018 Experimental definition of k-power of pre-exponential voltage factor for LAFE. Technical Digest, 2018 31st International Vacuum Nanoelectronics Conf. (IEEE, Piscataway, 2018), pp. 254-255.

23. Forbes R G. 2008 Call for experimental test of a revised mathematical form for empirical field emission current-voltage characteristics. *Appl. Phys. Lett.* **92**, 193105. (doi:10.1063/1.2918446)